# SPH SIMULATIONS OF REGULAR AND IRREGULAR WAVES AND THEIR COMPARISON WITH EXPERIMENTAL DATA

D. De Padova[1], R. A. Dalrymple[2], M. Mossa[1], A. F. Petrillo[3]

(1) Environmental Engineering and Sustainable Development Department, Technical University of Bari, Via E. Orabona, 4 – 70125 Bari, Italy, e-mail: d.depadova@poliba.it, m.mossa@poliba.it
(2) Department of Civil Engineering, Johns Hopkins University, 3400N Charles Street, Baltimore, MD 21218, USA, e-mail: rad@jhu.edu
(3) Water Engineering and Chemistry Department, Technical University of Bari, Via E. Orabona, 4 – 70125 Bari, Italy, e-mail: petrillo@poliba.it

**Abstract**

This paper presents the Smoothed particle hydrodynamics (SPH) model to examine the propagation of a regular and irregular waves.

The SPH method is a grid-less Lagrangian approach which is capable of good accuracy in tracking large deformations of a free surface.

The computations are validated against the experimental data and a good agreement is observed. The SPH modelling is shown to provide a promising tool in predicting the transformation of different waves.

## 1. INTRODUCTION

Smoothed particle hydrodynamics (SPH) is a meshfree, Lagrangian particle method for modeling fluid flows. SPH was first invented to solve astrophysical problems in three – dimensional open space, in particular polytropes (Lucy, 1977; Gingold and Monaghan, 1977), since the collective movement of those particles is similar to the movement of a liquid or gas flow, and it can be modeled by the governing equations of the classical Newtonian hydrodynamics.

Although the traditional grid – based numerical methods such as the finite difference methods (FDM) and the finite element methods (FEM) exist, and in some circumstances, are better developed methods than the SPH method, they have difficulties in handling some complex phenomena. This motivated researchers to seek for alternatives to solve these kinds of problems, and the SPH method has then become a good choice. Several review papers on the SPH method have been published, including those by Benz (1990) and Monaghan (1992).

These particles are capable of moving in the space, carry all the computational information, and thus form the computational frame for solving the partial differential equations describing the conservation laws of the continuum fluid dynamics.

The SPH method has been applied extensively to a vast range of problems in either computational fluid or solid mechanics because of relatively strong ability to incorporate complicated physical effects into the SPH formulations.

The application of SPH to a wide range of problems has led to significant extensions and improvements of the original SPH method. The numerical aspects have been gradually improved, some inherent drawbacks of SPH were identified, and modified techniques or corrective methods were also proposed.

Over the past years, different modifications or corrections have been tried to restore the consistency and to improve the accuracy of the SPH method. These modifications lead to various versions of the SPH methods and corresponding formulations. Currently, the SPH is a method that can simulate general fluid dynamic problems fairly well.

Though the SPH method has been extensively applied to different areas, there are still a lot of issue that need to be investigated further. This is particularly true is the numerical analyses of the method. There is still a long way to go for the method to become extensively applicable, practically useful and robust as the traditional grid – based methods such as FEM and FDM. This is because much work needs to be done to consolidate the theoretical foundations of SPH method, and to remedy its inherent numerical drawbacks. Moreover, there should be a necessary process for any numerical technique to develop, advance, improve, and to be validate so as to be lore efficient, robust in practical applications.

The development of the SPH method is still ongoing and the numerical model results require further analysis and detailed comparison with experimental data. The present paper presents the modelling of the propagation of regular and irregular breaking waves using the SPH approach.

In addition to theory, comparisons with physical model runs are analyzed, demonstrating the important role of the smoothing function in terms of computational accuracy. Moreover, an artificial viscosity is used, following Monaghan (1992). However, there were some difficulties in establishing the correct value of the fluid viscosity.

The empirical coefficient α, used in artificial viscosity (Monaghan, 1992), is needed for numerical stability in free-surface flows, but in practice it could be too dissipative.

The study made particular reference to the velocity and free surface elevation distributions. The final model is shown to be able to model the propagation of regular and irregular waves.

## 2. THEORETICAL BACKGROUND METHODOLOGY

The numerical technique (SPH) is a gridless, pure Lagrangian method for solving the equations of fluid dynamics. The main features of the SPH method, which is based on integral interpolations, were described in detail by Monaghan (1982), Benz (1990), Monaghan (1992) and Liu (2003). The alternative view is that the fluid domain is represented by nodal points that are scattered in space with no definable grid structure and move with the fluid. Each of these nodal points carry scalar information, density, pressure, velocity components and so on. To find the value of a particular quantity $f$ at an arbitrary point, $x$, we apply an interpolation:

$$f(x) = \sum_j f_j W(x - x_j) V_j \quad (1)$$

Here $f_j$ is the value of $f$ associated with particle $j$, located at $x_j$, $W(x-x_j)$ represents a weighting of the contribution of particle $j$ to the value of $f(x)$ at position $x$, and $V_j$ is the volume of particle $j$, defined as the mass, $m_j$, divided by the density of the particle $\rho_j$.

The smoothing function, $W(x-x_j)$, is called the Kernel and varies with the distance from $x$. The efficiency and accuracy of the constructed smoothing functions have also been shown in various literatures for all existing smoothing functions. Readers are referred to work by Fulk and Quinn (1995). The Kernel is assumed to have compact support, so the sum is only taken from neighboring particles. To some extent it should not matter which Kernel is used in SPH as long as basic requirements are met. This is especially true in the limits where $h$ (the Kernel smoothing length) and $\Delta x$ (the interparticle spacing) become small.

However, when these are not small, as is common in practice, the choice of Kernel can drastically change the computational results. Hence, the choice of Kernel, $h$ and $\Delta x$ is a key decision before performing any calculation using SPH. This paper provides an objective means of separating better from poorer Kernel performance in terms of $\Delta x/h$ value. In performing the analysis we consider the cubic spline Kernel and its first derivative. Monaghan and Lattanzio (1985) devised the following function based on the cubic spline function known as the B-spline function:

$$W(R,h) = \alpha_d \times \begin{cases} \frac{2}{3} - R^2 + \frac{1}{2}R^3 & 0 \leq R < 1 \\ \frac{1}{6}(2-R)^3 & 1 \leq R < 2 \\ 0 & R \geq 2 \end{cases} \qquad (2)$$

where $R = r_{ij}/h$, $r_{ij}=|x_i-x_j|$ and $\alpha_d =1/h$ for 1D, $\alpha_d =15/7\pi h^2$ for 2D, and $\alpha_d =3/2\pi h^2$ for 3D.

The cubic spline function has so far been the most widely used smoothing function in emerging SPH literature, since it resembles a Gaussian function while having a narrower compact support.

In smoothed particle hydrodynamics, the fluid has been traditionally considered compressible. The reason is that it is easier to calculate the pressure from an equation of the state rather than having to solve for the pressure. The conservation of mass and the conservation of the moment are written in particle form (Monaghan, 1992)

$$\frac{d\rho_i}{dt} = \sum_j m_j (u_i - u_j) \cdot \nabla_i W_{ij} \qquad (3)$$

$$\frac{du_i}{dt} = -\sum_j m_j \left( \frac{P_i}{P_i^2} + \frac{P_j}{P_j^2} + \Pi_{ij} \right) \cdot \nabla_i W_{ij} + g \qquad (4)$$

where $u_j$ is the velocity of the particle, $P_j$ is the pressure at the particle, $m_j$ is the mass of the particle $j$, $\nabla_i W_{ij} = \nabla_i W(x_i - x_j) = \frac{\partial W_{ij}}{\partial x_i}\mathbf{i} + \frac{\partial W_{ij}}{\partial y_i}\mathbf{j}$, $\Pi_{ij}$ is an empirical term representing the effects of viscosity (Monaghan, 1992):

$$\Pi_{ij} = -\frac{\alpha \mu_{ij} \bar{c}_{ij}}{\bar{\rho}_{ij}} \qquad (5)$$

where $\alpha$ is an empirical coefficient (usually taken as 0.01-0.1), $\bar{c}_{ij} = (c_i + c_j)/2$ and $\mu_{ij} = h(u_i - u_j)\cdot(x_i - x_j)/(r_{ij}^2 + 0.01h^2)$. Particles are moved using the XSPH variant due to Monaghan (1989):

$$\frac{dr_i}{dt} = u_i + \varepsilon \sum_j m_j \left( \frac{u_i - u_j}{\bar{\rho}_i} \right) \cdot W_{ij} \qquad (6)$$

where $\varepsilon = 0.5$ is the so-called XSPH correction of Monaghan (1989), which ensures that neighboring particles move with approximately the same velocity. This prevents particles with different velocities occupying nearly the same location.

The fluid was treated as weakly compressible in the present approach. The Weakly Compressible SPH (WCSPH) method has been utilized for many applications such as runup and rundown of waves on beaches (Monaghan and Kos, 1999), wave breaking on arbitrary structures (Colagrossi and Landrini, 2003) and wave breaking (Dalrymple and Rogers, 2006). An incompressible SPH was developed by Lo and Shao (2002).

Following Batchelor (1974), the relationship between pressure and density was assumed to follow the expression:

$$P = B\left[\left(\frac{\rho}{\rho_0}\right)^\gamma - 1\right] \qquad (7)$$

where $\gamma = 7$ and $B = c_0^2 \rho_0 / \gamma$, being $\rho_0 = 1000$ $Kg\ m^{-3}$ the reference density and $c_0 = c(\rho_0) = 465.9$ $ms^{-1}$ the speed of sound at the reference density. This sound speed is lower than that of a real fluid, but much faster than any waves in the model.

The advantage of using an equation of state is that there is no need to solve for a partial differential equation for pressure, which is time consuming to solve.

An incompressible SPH was developed by Lo and Shao (2002). The advantage of this model is that the fluid is incompressible as in the Moving Particle Semi-implicit (MPS) method (Koshizuka et al., 1995) and the time step is based on the maximum fluid velocity, not the sound speed; the disadvantage is that the pressure has to be numerically obtained, requiring many more calculations. A question that is currently be solved is whether the larger time steps but more calculation is faster than the SPH with its smaller time steps but with a simple pressure calculation (the equation of state).

A detailed comparison between the Weakly compressible SPH and the Incompressible SPH can be found in Shipilova et al. (2009).

We have no experience in carrying out the Incompressible SPH computations, and therefore, a direct comparison of the CPU times and numerical accuracies between the two kinds of SPH approaches are not possible in this paper.

**3. MODEL VERIFICATION**

To test performance of the SPH model, numerical results were compared with experimental results of regular and irregular breaking waves. Comparisons began with a regular wave test on a plane slope (De Serio & Mossa (2006)). Next, a run of irregular wave test on an irregular slope bottom, was simulated. In addition to velocity, time series of surface elevation and a wide range of statistical parameters were compared for these tests.

**3.1 Experimental set-up**

**Regular wave**

The experiments were performed in a wave channel 45 m long and 1 m wide. The iron frames supporting its crystal walls are numbered from the shoreline up to the wavemaker (section 100), thus locating measurement sections which have a center to center distance equal to 0.44 m. From the wave paddle to section 73 the flume has a flat bottom, while from section 73 up to the shoreline it has a 1/20 sloped wooden bottom.

A sketch of the wave flume is shown in Fig. 1. Further details about the experimental tests carried out can be found in De Serio and Mossa (2006).

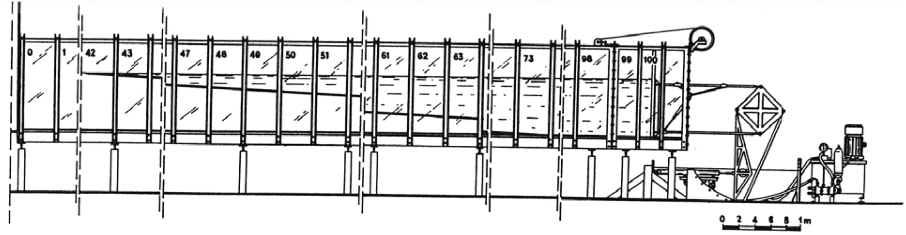

Figure 1. Sketch of the wave flume.

Table 1 shows the main characteristics of the tested regular wave with a wave period T= 2s, evaluated in different sections, where $d$ is the mean water depth, $H$ the wave height and period, $L_A$ the wave length according to the Airy theory, $U=HL_A^2/d^3$ the Ursell number, indicator of a non-linear wave behaviour. Taking into account the Irribarren breaking number $\xi_b$, the wave fields of the present study are characterized by a spilling/plunging breaker in the test ($\xi_b$=0.37).

Like most of the recent experimental investigations (Okaiasu et al., 1986; Feng and Stansby, 2002), the Laser Doppler Anemometry, a modern non-intrusive measurement technique, was used to measure the instantaneous Eulerian velocity. In particular, a backscatter, two-component four beam LDA system and a Dantec LDA signal processor (58N40 FVA Enhanced) based on the covariance technique was adopted. The wave elevations were measured with a resistance probe placed in the transversal section of the channel crossing the laser measuring volume. The velocity and wave elevation measurements were assessed simultaneously, allowing us to perform the phase-averaging analysis. The velocity components measured in the present study are u in the x direction, conventionally established as positive if oriented onshore, and v in the vertical direction, conventionally established as positive if oriented upward. All the velocity measurements were assessed in many vertical sections between the bottom boundary layer and the wave trough, i.e. in the middle region of surf-zone, due to the fact that LDA suffers from signal drop-out within the aerated region

| Section | d[cm] | H[cm] | $L_A$[m] | U | Zone |
|---|---|---|---|---|---|
| 76 | 70.0 | 11.0 | 4.62 | 6,8 | Shoaling |
| 55 | 31.0 | 12.4 | 3.31 | 46 | Shoaling |
| 49 | 16.5 | 14.3 | 2.47 | 194 | Shoaling |
| 47 | 11.3 | 5.9 | 2.06 | 174 | Outer |
| 45 | 8.5 | 4.9 | 1.80 | 259 | Inner |

Table 1: Principal characteristics of tasted wave

.
**Irregular Wave**

The study of a random wave is of greater practical interest, because most of the waves found in the nature are complex, consisting of numerous superimposed components of the wave period, the wave height and wave direction. The regular waves are seldom found in the field and thus are increasingly less frequently used in the laboratory experiments.

A series of 2D experiments were carried out in a large mobile-bed channel of the Coastal Engineering Laboratory (LIC) of the Technical University of Bari.

The wave channel at the LIC is 50 m long, 2.4 m wide and 1.2 m deep (Figure 2 and 3). The channel is presently equipped with a 2D piston-type wave generator formed by 1 modules 0.6 m long, for a total length of 2.4 m, capable of generating waves up to 30 cm high. Irregular long-crested waves were generated with the standard software (HR Wallingford). JONSWAP spectral shape was used for the random waves.

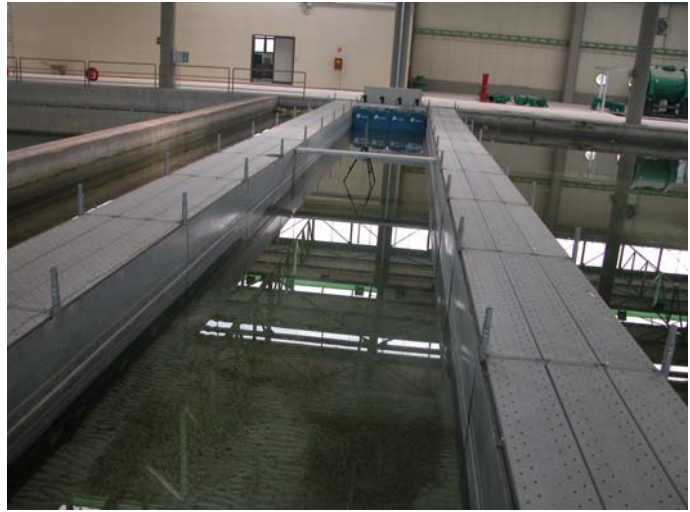

Figure 2: Sketch of the wave flume.

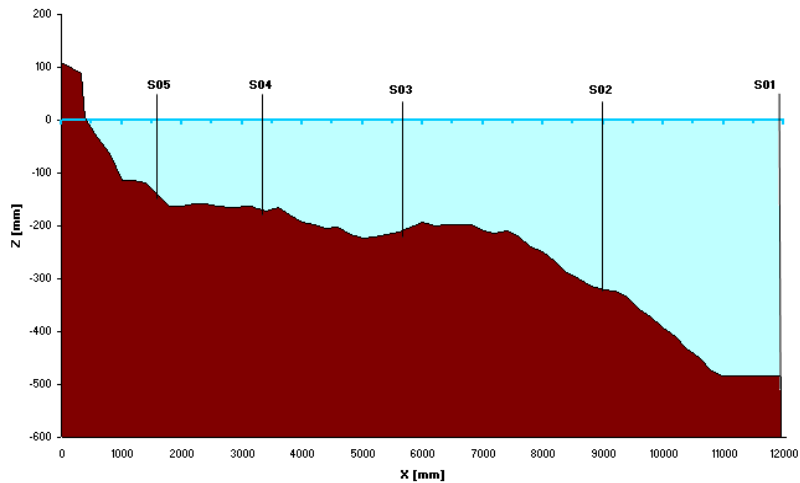

Figure 3: Measurement sections.

Table 2 shows the random wave parameters of the three tested random waves, such as the significant wave height, $H_s$, the peak wave period, $T_p$, where $d$ is the mean water depth.

|  | Section | d [cm] | $H_s$ [cm] | $T_p$ [s] |
|---|---|---|---|---|
| Test1 | SO1 | 47.7 | 5.0 | 1.78 |
| Test2 | SO1 | 47.7 | 7.6 | 1.5 |
| Test3 | SO1 | 47.7 | 15.7 | 1.78 |

Table 2: Principal characteristics of the laboratory random wave tests

The importance of field measurements is highlighted in order to calibrate and quantitatively validate the numerical models results in design applications using the SPH code.

**3.2 Numerical test**

**3.2.1 Regular wave**

The computational wave tank is taken to be 22.5 m long and 0.97 m high, which is shorter than the experimental one. This choice of a shorter tank only reduced the computational cost. The choice of the Δ$x/h$ term depends on the physical process of the problem and the desired computational accuracy and efficiency. However, if an interval of Δ$x/h$ value is studied, the quality of Kernel as particle movement can be deduced. As shown in Fulk and Quinn (1995), for almost every Kernel, the results start to become relatively poor when Δ$x=h$. However, we verified this result for SPH calculations. In a previous study, it was studied that the configuration with 30,000 particles is the best fit to the data (De Padova et al., 2008). The particle spacing is taken as Δ$x$=Δ$z$=0.022 m and thus approximately 30,000 particles are used. For a comparison between computational accuracies, we used a smoothing length of $h$=0.0305 m and $h$=0.0212 m and, thus, the values of Δ$x/h$=0.7213 (Test 1) and Δ$x/h$=1.0377 (Test 2) were used.

The first two simulations in the present paper used an artificial viscosity with an empirical coefficient α of equation 5, equal to 0.055 (Monaghan, 1992).

Another simulation was carried out (Test 3), to study the effects of fluid viscosity in terms of stability and dissipation in the fluid calibrating the empirical coefficient α. In this simulation (Test 3) we used a value of the empirical coefficient α equal to 0.045.

In all simulations, the water depth, the wave height and the period were equal to 0.70 m, 0.11 m and 2 s, respectively, in section 0.5 m offshore the section 76 (Figure 4).

During the several SPH computations, the time step Δ$t$ was adjusted to satisfy the stability requirement by Courant condition and viscous term constrain (Monaghan,1992).

The Predictor-Corrector algorithm (Monaghan, 1989) with variable time step (Monaghan and Kos, 1999), was applied in the numerical simulations.

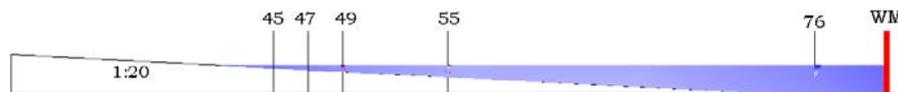

Figure 4: Measurement sections.

The walls of the tank were built with two parallel layers of fixed boundary particles placed in a staggered manner described by Dalrymple and Knio (2000). In this approach the boundary particles share some of the properties of fluid particles, but their velocities are zero and their positions remain unchanged.

There is a computational advantage in this treatment of the boundaries as fixed fluid particles (Dalrymple and Knio, 2000); it is not necessary to make an a priori assumption about the nature of the force exerted by the boundaries.

The current incompressible SPH computations are executed up to t=25s from the beginning and it costs about 40 hours of CPU time using a personal computer (CPU 2.66 GHz and RAM 4.0 GMB PC). Table 3 shows the main characteristics of the SPH simulation.

| Test | Boundary conditions | Time simulation | Time stepping scheme | Coefficient in the artificial viscosity (α) | Particle number | $\Delta x/h$ |
|---|---|---|---|---|---|---|
| 1 | Repulsion boundary of Dalrymple and Knio (2000) | 25s | Predictor corrector scheme $(\Delta t = 1\times 10^{-4} s)$ | 0.055 | 30,000 | 0.7213 |
| 2 | Repulsion boundary of Dalrymple and Knio (2000) | 25s | Predictor corrector scheme $(\Delta t = 1\times 10^{-4} s)$ | 0.055 | 30,000 | 1.0377 |
| 3 | Repulsion boundary of Dalrymple and Knio (2000) | 25s | Predictor corrector scheme $(\Delta t = 1\times 10^{-4} s)$ | 0.045 | 30,000 | 0.7213 |

Table 3: Principal characteristics of SPH simulations

The study made particular reference to analyze the performance of Kernel in terms of stability in the fluid with the first two tests and to calibrate the empirical coefficient α, used in the artificial viscosity with the test 3.

### 3.2.1.1 Results

The experimental and numerical wave profiles at the location of measurement points are shown for all three cases. For a defined section, we can study the distribution along the channel of the wave elevation and the horizontal and vertical velocity components for all three tests. Figure 5a÷5c, 6a÷6c, 7a÷7c, 8a÷8c and 9a÷9c show the agreement of numerical data obtained by means of the first two SPH models with experimental data (Table 3).

In Figs 5a÷5c, 6a÷6c, 7a÷7c, 8a÷8c and 9a÷9c it can be seen that the numerical elevations and the numerical velocities obtained by means of the second test of table 3, are not in perfect agreement with the experimental measurements for the strong effect of the Δx/h term. In fact, when this value becomes equal to 1 (Test 2), the computational results can drastically change for the worse. With a smaller value of the Δx/h term (Test 1), numerical elevations and the numerical velocities are shown to be in better agreement with the experimental measurements (Figs. 5a÷5c, 6a÷6c, 7a÷7c, 8a÷8c and 9a÷9c).

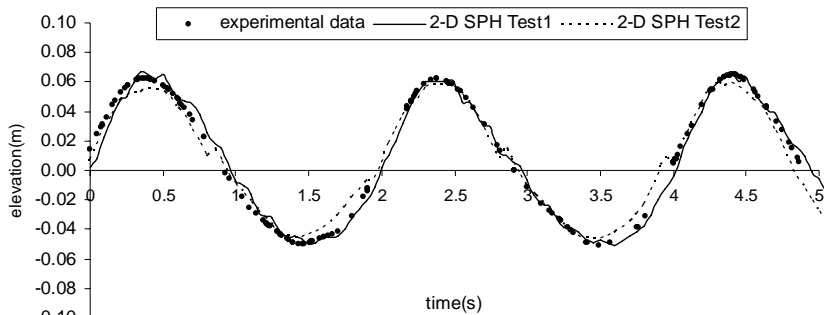

**Figure 5a:** Comparison of 2-D SPH with experimental data (section 76)

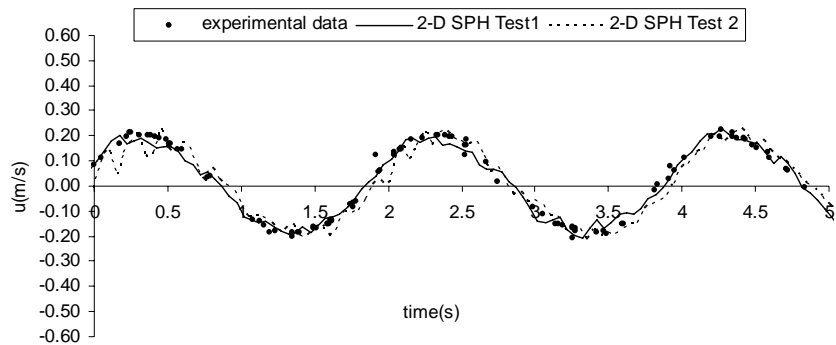

**Figure 5b**: Comparison of experimental and numerical horizontal velocity components (section 76, 0.33 m from the bottom)

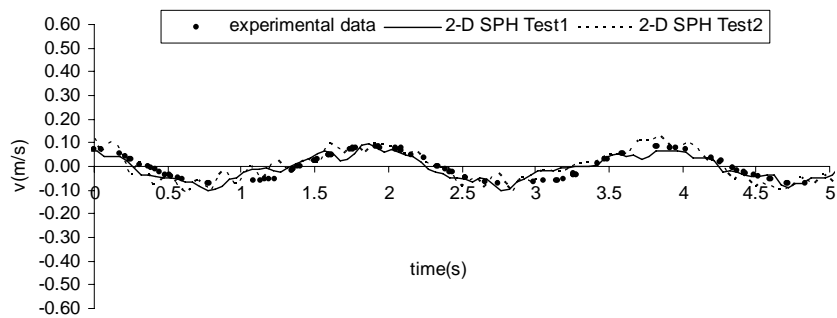

**Figure 5c**: Comparison of experimental and numerical vertical velocity components (section 76, 0.33 m from the bottom)

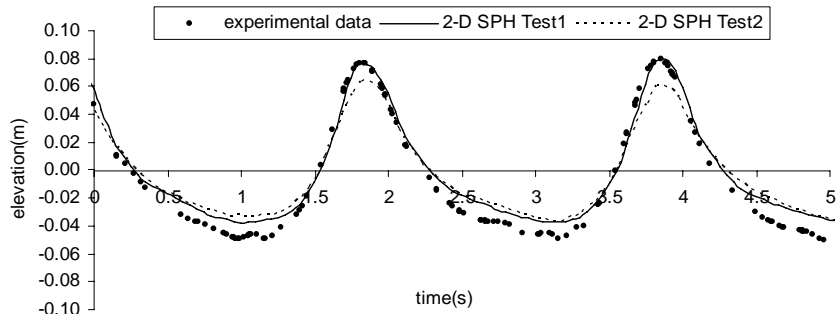

**Figure 6a:** Comparison of 2-D SPH with experimental data (section 55)

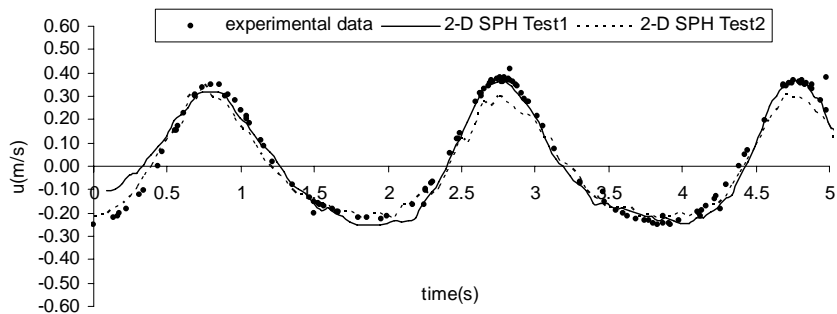

**Figure 6b:** Comparison of experimental and numerical horizontal velocity components (section 55, 0.1 m from the bottom)

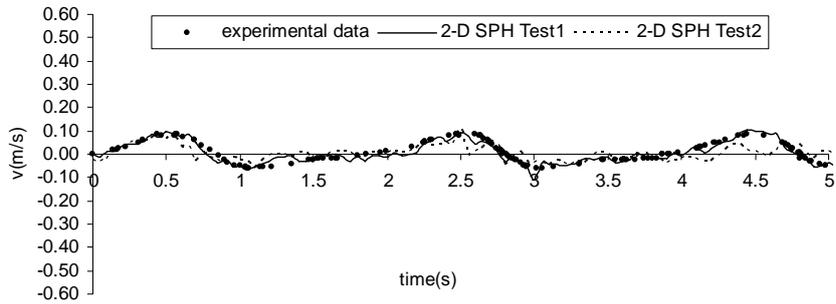

**Figure 6c:** Comparison of experimental and numerical vertical velocity components (section 55, 0.1 m from the bottom)

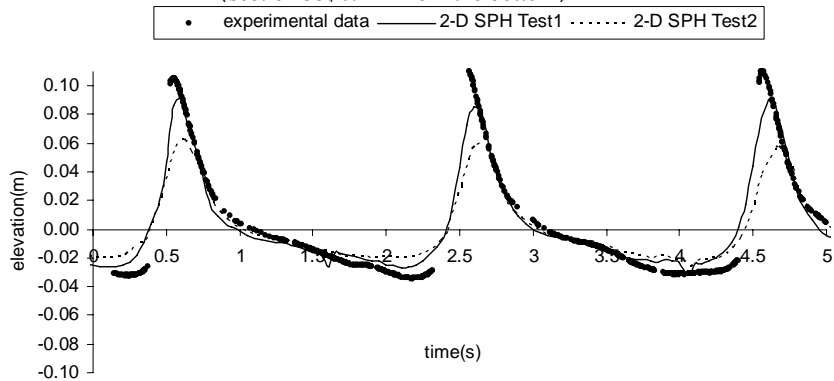

**Figure 7a:** Comparison of 2-D SPH with experimental data (section 49)

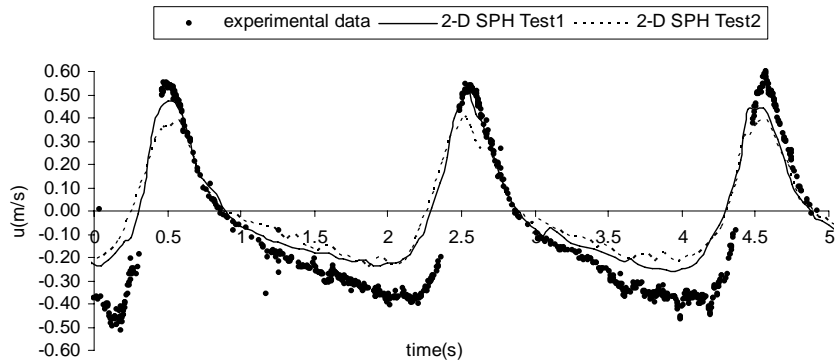

**Figure 7b:** Comparison of experimental and numerical horizontal velocity components (section 49, 0.1 m from the bottom)

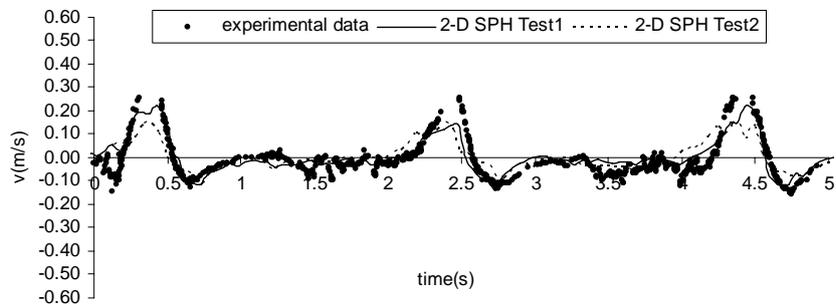

**Figure 7c:** Comparison of experimental and numerical vertical velocity components (section 49, 0.1 m from the bottom)

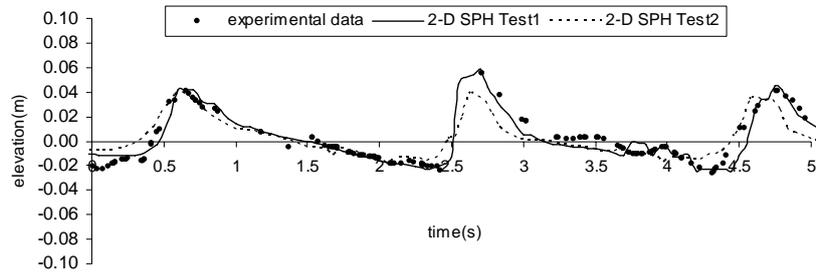
**Figure 8a:** Comparison of 2-D SPH with experimental data (section 47)

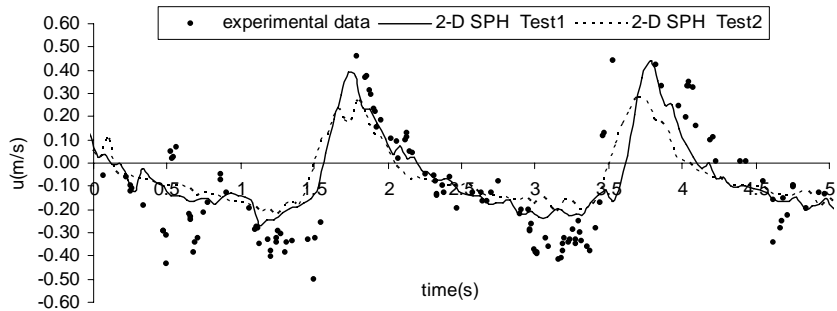
**Figure 8b:** Comparison of experimental and numerical horizontal velocity components
(section 47, 0.063 m from the bottom)

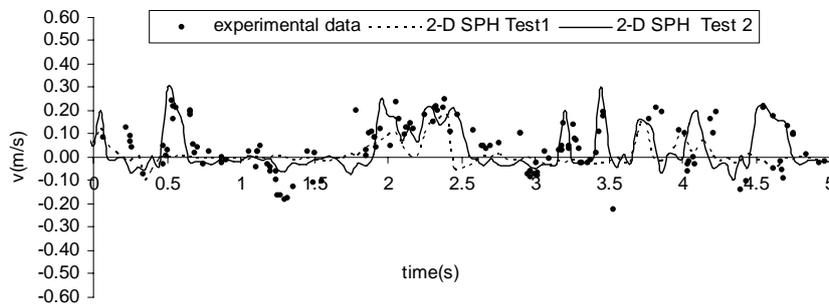
**Figure 8c:** Comparison of experimental and numerical vertical velocity components
(section 47, 0.063 m from the bottom)

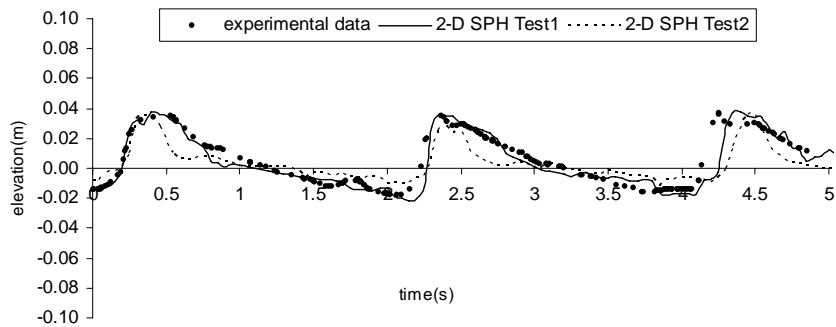
**Figure 9a:** Comparison of 2-D SPH with experimental data (section 45)

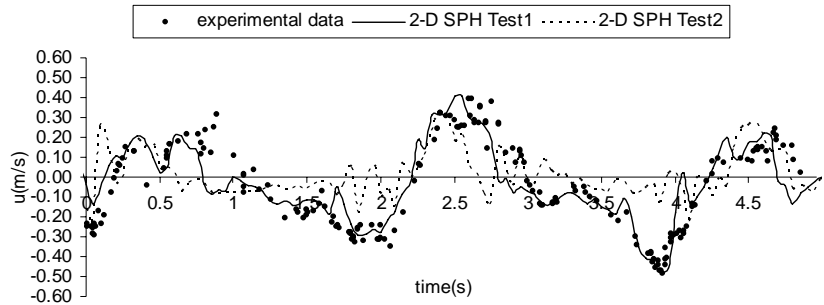

**Figure 9b:** Comparison of experimental and numerical horizontal velocity components
(section 45, 0.045 m from the bottom)

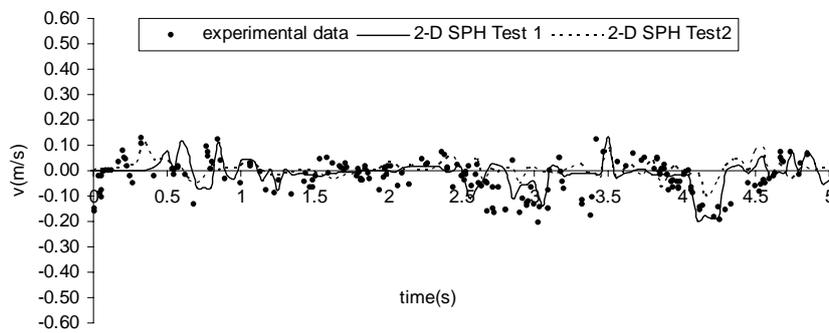

**Figure 9c:** Comparison of experimental and numerical vertical velocity components
(section 45, 0.045 m from the bottom)

Although Figs 5a÷5c, 6a÷6c, 7a÷7c, 8a÷8c and 9a÷9c provide good qualitative results, it is desirable to obtain quantitative results as well. Overall statistical parameters can provide a more detailed picture of the breaking model performance. Figure 10 shows standard deviations of measured and computed surface elevations of the sections 76, 55, 49, 47 and 45. Skewness (Kennedy et al., 2000), a measure of crest-trough shape, are computed and shown in Fig 11. The Test1 predicts this parameter very well; in fact the trend of wave skewness increases as the wave shoals and breaks, and decreasing near the shoreline (section 49). Instead, in the case of Test 2 of Table 3, when the value of $\Delta x/h$ term becomes equal to 1, the computational results change for the worse and the trend of wave skewness is not predicted well in particular at sections 47 and 45 where wave profile is characterized by a rapid change in shape. These result shows how a good efficiency of the Smoothed Particle Hydrodynamics Kernel is not obtained for all values of $\Delta x/h$.

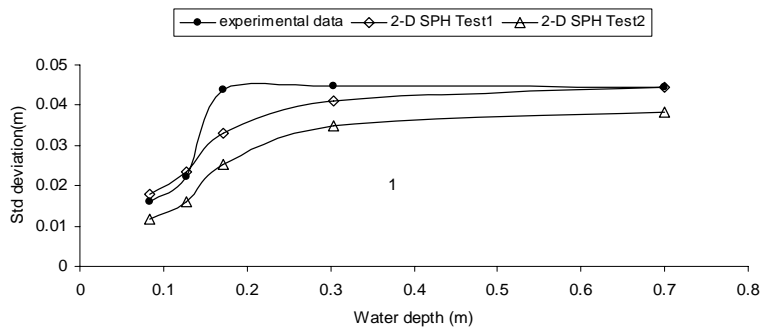

**Figure 10:** Comparison of experimental and numerical standard deviation of surface wave elevations

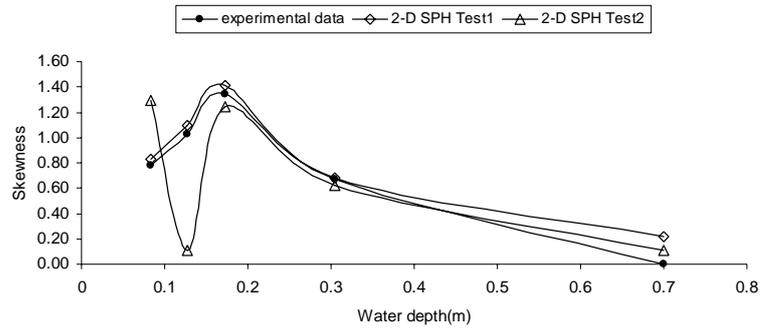

**Figure 11:** Comparison of experimental and numerical skewness of surface wave elevations

In Figs 6a and 7a it can be seen that the artificial viscosity seems to bring a relevant dissipative effect on the wave height before and during the breaking. Figures 12 and 13 show the agreement of numerical data obtained by means of the third SPH model with experimental data (Table 3). With a smaller empirical coefficient α (test 3), numerical elevations are shown to be in better agreement with the experimental measurements (Figg. 12 ÷ 13). These results show how the empirical coefficient α is needed for numerical stability for free- surface flows, but in practice it is too dissipative and an appropriate and calibrated value of α should be used for a settled configuration in order to obtain good results.

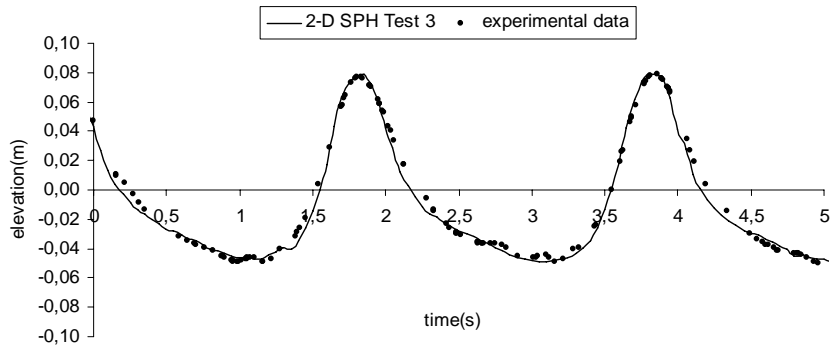

**Figure 12:** Comparison of 2-D SPH with experimental data (section 55)

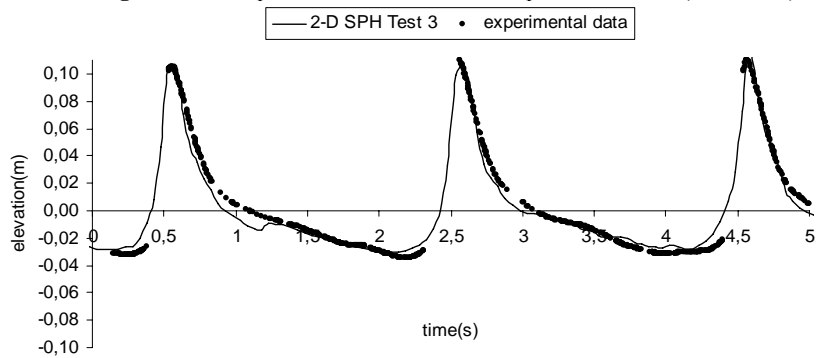

**Figure 13:** Comparison of 2-D SPH with experimental data (section 49)

### 3.2.2 Irregular wave

To test model SPH performance using an irregular wave, the third experimental test was simulated with the worst condition (Table 2). Computation was initialized using a time series of data at the deepest measurement location, denoted SO1 in Figure 3.

The computational wave tank is taken to be 12.5 m long and 1.2 m high, which is shorter than the experimental one. As previously written for the regular wave, this choice of a shorter tank only reduced the computational cost. Particles were initially placed on a staggered grid with zero initial velocity (Figure 14). The boundaries were modelled employing Monaghan's repulsive boundary conditions with wall viscosity value for Repulsive Force BC equal to 2.0e-4 (Monaghan and Kos, 1999).

The initial particle spacing is taken to be $\Delta x=\Delta z=0.022$ m and thus approximately 13,000 particles are used. Once again, a smoothing length of $h=0.0305$ was used and consequently the previously studied value of $\Delta x/h=0.7213$ was also adopted.

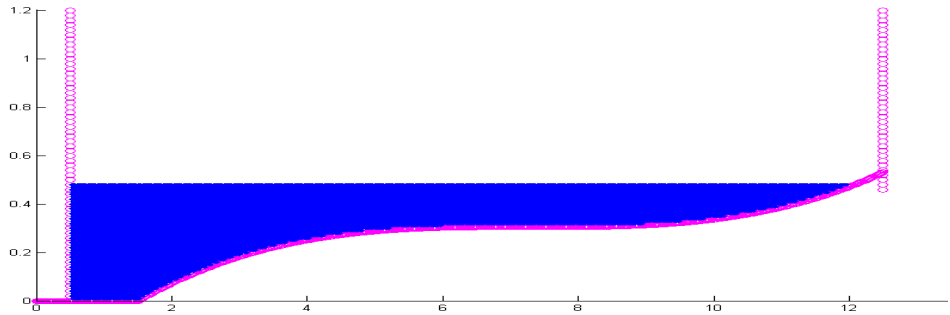

Figure 14: Initial smoothed particle hydrodynamics configuration of fluid and boundary

The simulation uses an artificial viscosity with a calibrated empirical coefficient $\alpha$ of equation 5, equal to 0.035 (Monaghan, 1992). The water depth was equal to 0.477 m in section 0.5 m offshore the section SO1 (Figure 14).

The random wave trains are generated by the superposition of a finite number of linear wave modes with different wave heights and wave frequencies.

During the several SPH computations, the time step $\Delta t$ was adjusted to satisfy the stability requirement by Courant condition and viscous term constrain (Monaghan,1992). A single predictor correct scheme similar to that described in Monaghan (1989) was used. A constant time step $\Delta t = 1 \times 10^{-4}$s was used in the all calculations. The SPH model is run for 120s and the computations are finished within 60 h by using a CPU 2.66 GHz and RAM 4.0 GMB PC.

| Test | Boundary conditions | Time simulation | Time stepping scheme | viscosity | N° Particles |
|---|---|---|---|---|---|
| 1 | Monaghan's repulsion boundary | 120s | predictor corrector scheme $\left(\Delta t = 1\times 10^{-4} s\right)$ | Artificial viscosity $\alpha = 0.035$ | 13,000 |

Table 5.1: Characteristics of SPH simulation

The study made particular reference to the free surface elevation distributions with the aim of analysing the performance of SPH model simulating an irregular wave. The final model proved capable of reproducing the experimental propagation of irregular and breaking wave.

For a defined section, we can study the distribution of the wave elevation for the test along the channel. Figures 15a÷15b 16a÷16b 17a÷17b 18a÷18b show a time series of computed and measured surface elevations and the spectrum at the other four sections, SO2, SO3, SO4 and SO5 (Figure 2). The spectra were arrested at 1 Hz for the sake of clarity. Agreement is in general quite good, but deteriorates as the wave progress toward shore.

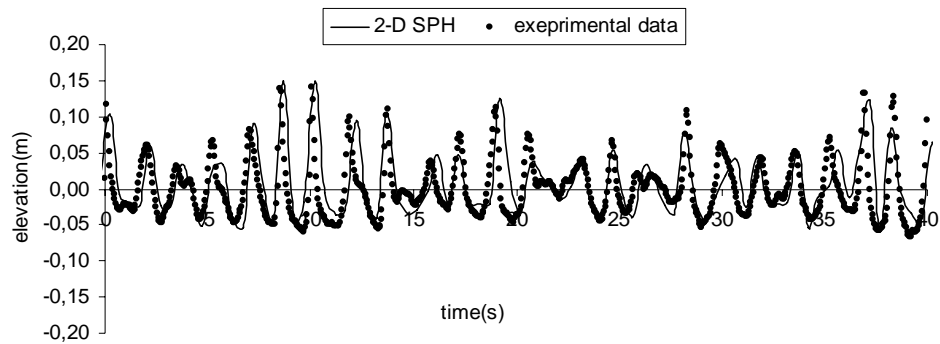

Figure 15a: Comparison of 2-D SPH with the first 40s of the experimental data (section S02)

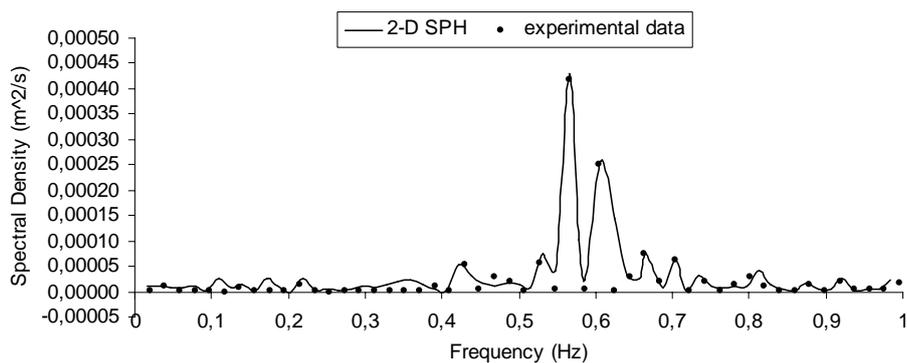

Figure 15b: Comparison of 2-D SPH with experimental data (section S02)

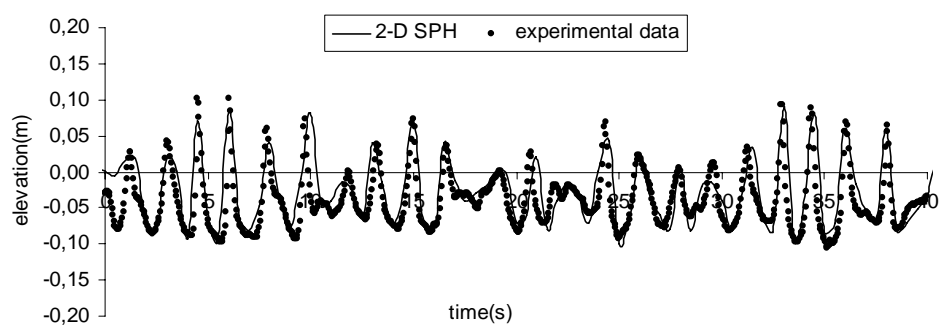

Figure 16a: Comparison of 2-D SPH with with the first 40 s of experimental data (section S03)

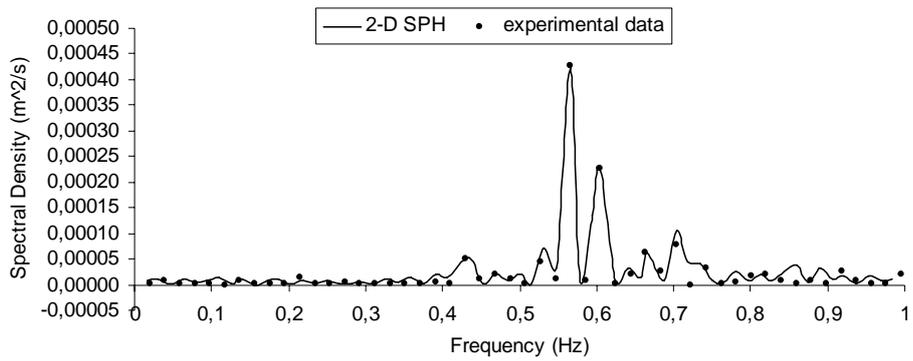

Figure 16b: Comparison of 2-D SPH with experimental data (section S03)

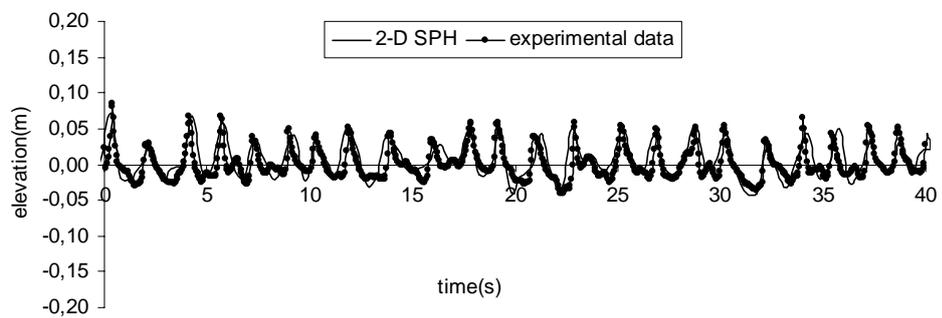

Figure 17a: Comparison of 2-D SPH with the first 40 s of experimental data (section S04)

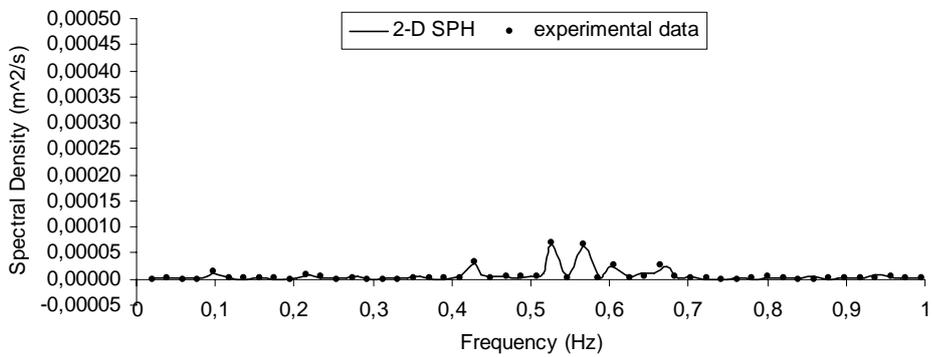

Figure 17b: Comparison of 2-D SPH with experimental data (section S04)

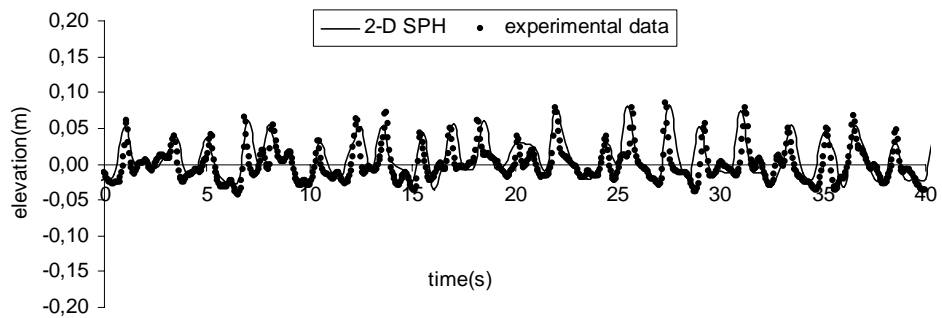

Figure 18a: Comparison of 2-D SPH with the first 40 s experimental data (section S05)

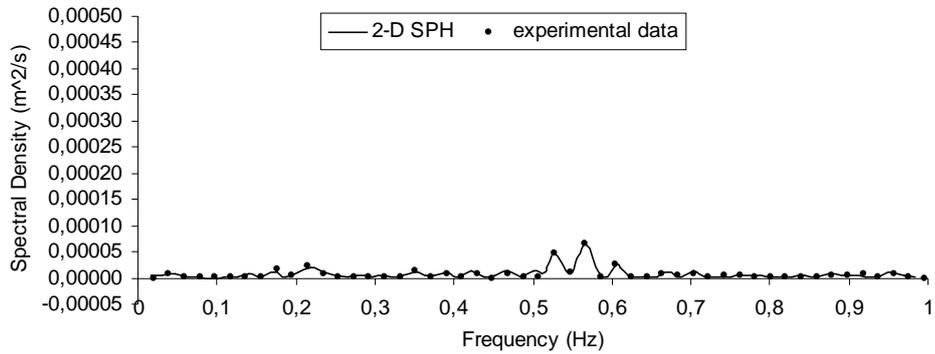

Figure 18b: Comparison of 2-D SPH with experimental data (section S05)

Overall statistical parameters can provide a more detailed picture of the breaking model performance. Figure 19 shows standard deviations of measured and computed surface elevations of sections S02, SO3, S04 and S05 (Fig. 3). Skewness (Kennedy et al., 2000), a measure of crest-trough shape, is computed and shown in Fig. 20. The Test predicts this parameter very well; In fact the trend of wave skewness increases as the wave shoals and breaks, and decreases near the shoreline (section S03). Positive skewness here corresponds to narrow crests and flat thoughs.

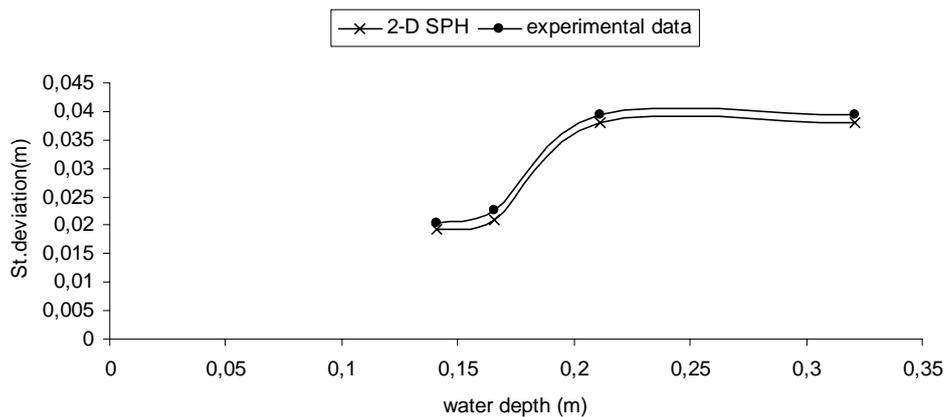

Figure 19: Comparison of experimental and numerical standard deviation of surface wave elevations

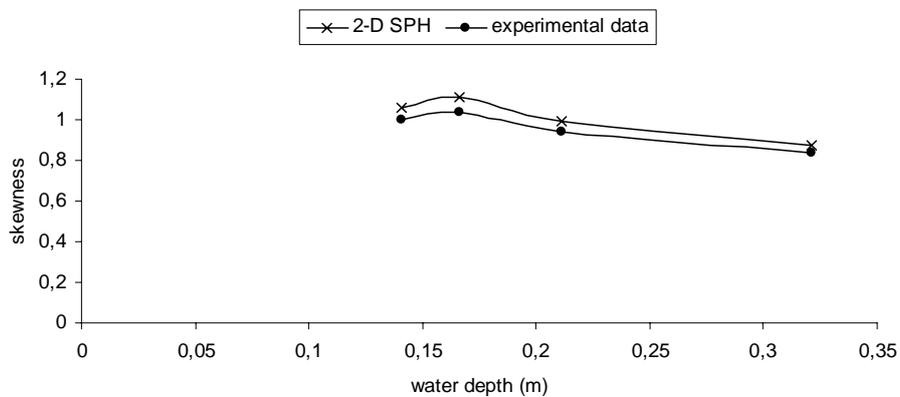

Figure 20: Comparison of experimental and numerical akewness of surface wave elevations

**Conclusion**

Numerical tests have been carried out to examine the ability and efficiency of the presented SPH formulations and the SPH code implemented in simulating fluid dynamic problems. The SPH model is employed to reproduce the propagation of regular and random breaking waves in a computational tank with a regular and irregular bottom's slope.

The computations are in general quite good agreement with the experimental data. The model has shown itself to accurately predict wave transformation in the surf and swash zone. Breaking phenomena are predicted both qualitatively and quantitatively for both cases runs demonstrating the potentialities of SPH like an engineering tool.

Turbulence at the wave breaking is treated by using an artificial viscosity. Several improvements that we have made in the same numbers of particles with different value of empirical coefficient α are presented here with the aim to study the effects of fluid viscosity in terms of stability and dissipation in the fluid. The importance of field measurements is highlighted in order to calibrate and quantitatively validate the SPH numerical model. The choice of the empirical coefficient α depends on the physical process of the problem and the desired computational accuracy and efficiency. These results show how the empirical coefficient α is needed for numerical stability for free - surface flows, but in practice it is too dissipative. Therefore, generally speaking, an appropriate and calibrated value of the empirical coefficient α should be used for a settled number of particle in order to obtain good results. By simulating a breaking waves, it is also shown that the efficiency of the SPH Kernel depends on the choice of the Δx/h term. These results highlight the fact that, for a certain Smoothed Particle Hydrodynamics Kernel, it is important to define and use a correct value of the Δx/h term in the model. This is particularly useful in selecting the initial particle separation for a given Kernel.